\documentclass[10pt, conference, compsocconf]{IEEEtran}
\usepackage{cite}
\usepackage[pdftex]{graphicx}
\usepackage{array}
\usepackage{url}
\usepackage{orcidlink}
\usepackage{colortbl}
\usepackage{amssymb}
\usepackage{algorithm}
\usepackage{fancyhdr}
\usepackage[noend]{algpseudocode}
\definecolor{LightCyan}{rgb}{0.4,0.8,1}
\definecolor{Gray2}{gray}{0.9}
\definecolor{backcolour}{rgb}{0.95,0.95,0.92}
\usepackage{listings}
\fancypagestyle{FirstPage}{
\lfoot{© 2023 IEEE.  Personal use of this material is permitted.  Permission from IEEE must be obtained for all other uses, in any current or future media, including reprinting/republishing this material for advertising or promotional purposes, creating new collective works, for resale or redistribution to servers or lists, or reuse of any copyrighted component of this work in other works.} 
}

\begin{document}

\title{Acceleration of a production Solar MHD code with Fortran standard parallelism: \newline From OpenACC to `do concurrent'}

\author{
\IEEEauthorblockN{
Ronald M. Caplan\,\orcidlink{0000-0002-2633-4290}, 
Miko M. Stulajter\,\orcidlink{0000-0003-0939-1055}, 
and Jon A. Linker\,\orcidlink{0000-0003-1662-3328}
}
\IEEEauthorblockA{
Predictive Science Inc. \\
9990 Mesa Rim Road, Suite 170 \\
San Diego, CA  92121 \\
caplanr@predsci.com}
}

\maketitle

\begin{abstract}
There is growing interest in using standard language constructs for accelerated computing, avoiding the need for (often vendor-specific) external APIs. These constructs hold the potential to be more portable and much more `future-proof'.  For Fortran codes, the current focus is on the {\tt do concurrent} (DC) loop.  While there have been some successful examples of GPU-acceleration using DC for benchmark and/or small codes, its widespread adoption will require demonstrations of its use in full-size applications.  Here, we look at the current capabilities and performance of using DC in a production application called Magnetohydrodynamic Algorithm outside a Sphere (MAS). MAS is a state-of-the-art model for studying coronal and heliospheric dynamics, is over 70,000 lines long, and has previously been ported to GPUs using MPI+OpenACC.  We attempt to eliminate as many of its OpenACC directives as possible in favor of DC.  We show that using the NVIDIA {\tt nvfortran} compiler's Fortran 202X preview implementation, unified managed memory, and modified MPI launch methods, we can achieve GPU acceleration across multiple GPUs without using a single OpenACC directive.  However, doing so results in a slowdown between 1.25x and 3x.  We discuss what future improvements are needed to avoid this loss, and show how we can still retain close to the original code's performance while reducing the number of OpenACC directives by a factor of five.
\end{abstract}


\IEEEpeerreviewmaketitle

\section{Introduction}
\label{sec:intro}
\thispagestyle{FirstPage}

The use of accelerators (such as GPUs) has become ubiquitous in high-performance computing.  This is due to their power efficiency and compact performance (i.e. one accelerator can perform as well as multiple CPU compute nodes).  However, developing and/or porting codes to run on accelerators continues to be a challenge, especially while maintaining portability.  Due to the historical lack of direct support for accelerators in the base standard languages (such as C, C++, and Fortran), a plethora of (often vendor-specific) accelerator APIs, libraries, and language extensions have been created over the years, each with varying degrees of support, portability, and software level (e.g. high-level front ends to low-level direct device programming).  These include CUDA, RocM, OpenCL, SYCL, KOKOS, RAJA, DPC++, PROTO, OpenMP, and OpenACC \cite{babelstream,proto,openacc_book,openmp_book}.  For legacy codes, as well as for new codes wanting maximum portability, longevity, and code readability, directive-based solutions such as OpenMP and OpenACC have often been preferred over other options.  This is because the directives appear as special comments in the code, allowing it to continue to compile and run on previous hardware/software setups, as the directives will simply be ignored.  Even with the great feature set of directive-based programming models, they can still make source code difficult to read and require too much training for the domain scientists to be able to develop and modify the code without breaking acceleration (often leading to multiple code bases).  Also, the specification for the directive models can change frequently, and cross-vendor support can be unpredictable.  

Recently, the use of standard language parallelism for accelerated computing has been gaining popularity.  This includes C++ parallel algorithms\cite{stdparcpp,cplusplusstdparnasa}, Fortran's {\tt do concurrent}\cite{stdpardc,inteldc}, and drop-in replacements for python's {\tt numpy}\footnote{\url{https://developer.nvidia.com/cunumeric}}.  These new language features have the potential to eliminate (or greatly limit) the need for external acceleration APIs, and could be more portable and `future-proof'.  Future portability is assured since any compiler that supports the base language version, must compile and run the code correctly (although not necessarily efficiently).  

Here, we focus exclusively on Fortran and its {\tt do concurrent} (DC) construct for GPU acceleration. DC is an alternative construct to the standard {\tt do} loop which indicates that the loop has no data dependencies and can be computed out-of-order\footnote{\url{https://flang.llvm.org/docs/DoConcurrent.html}}.  This can be used to hint to a compiler that the loop is likely (but may not be) parallelizable.  How this parallelism is mapped at a low level is left to the compiler to decide, with many using pre-existing directive-based mappings (treating DC as if it were a collapsed OpenACC/MP parallel loop).

 While there have been some promising results in using DC for accelerated computing (see the next section), how well they will extend to much larger codes is an open question.  To help answer this, in this paper we describe porting our large in-production solar MHD code from OpenACC to DC.

The paper is organized as follows:  Sec.~\ref{sec:relwork} describes related work on using Fortran standard parallelism for accelerated computing.  We then introduce the MAS solar MHD code in Sec.~\ref{sec:mas_intro}.  In Sec.~\ref{sec:acc2dc} we describe how we successively reduce the number of OpenACC directives using DC and the resulting six code versions used for testing.  The test problem, computational environment, and performance results are described in Sec.~\ref{sec:perf}.  We summarize the results and use them to assess the current status and future potential of Fortran standard parallelism for large HPC codes in Sec.~\ref{sec_summary}.

\section{Related Work}
\label{sec:relwork}

The use of DC for accelerated computing is a very recent capability. The first compiler to support it was NVIDIA's HPC SDK in November of 2020 \footnote{\url{https://developer.nvidia.com/blog/}\\ \url{accelerating-fortran-do-concurrent-with-gpus-and-the-nvidia-hpc-sdk/}}, and the most recent is the Intel IFX compiler in 2022.

Due to the novelty of DC support for GPUs, there are only a few examples of its use in the literature (such as a spherical surface diffusion tool called DIFFUSE \cite{stdpar_dc_waccpd}, and an implementation of the BabelStream benchmark\cite{BabelStream:Fortran}).    There are additional works in progress, including a chemistry mini-app called CCSD(T)\footnote{\url{https://www.youtube.com/watch?v=DrvI2gw3tnI}}, a CFD weather mini-app called MiniWeather\footnote{\url{https://github.com/mrnorman/miniWeather/tree/main/fortran}}, a hydrodynamics mini-app called CloverLeaf \footnote{\url{https://github.com/UoB-HPC/cloverleaf_doconcurrent}}, and a conjugate gradient solver used in Solar physics called POT3D\footnote{\url{https://www.nvidia.com/en-us/on-demand/session/gtcspring22-s41318}}.

While these implementations are promising (and many yielded on-par performance with other acceleration methods), they are all small-to-medium codes/mini-apps/benchmarks.  Here, we explore applying the lessons learned from these first studies to a large, in-production application called MAS.

\section{The MAS Solar MHD Model}
\label{sec:mas_intro}

The Magnetohydrodynamic Algorithm outside a Sphere (MAS) code is an in-production MHD model with over 20 years of ongoing development used extensively in Solar physics research \cite{mikic99a,2003PhPl...10.1971L,lionello06a,2009ApJ...690..902L,2011ApJ...731..110L,2013Sci...340.1196D,ECLIPSE2017}.  The code is included in the Corona-Heliosphere (CORHEL) software suite \cite{rileyetal2012,Feng2020} hosted at NASA's  Community Coordinated Modeling Center (CCMC)\footnote{\url{https://ccmc.gsfc.nasa.gov}} allowing users to generate quasi-steady-state MHD solutions of the corona and heliosphere, as well as simulate solar storms in the form of coronal mass ejections propagating from the Sun to Earth \cite{lionelloetal2013,torok18}. MAS is written in Fortran ($\approx 70,000$ lines) and parallelized with MPI+OpenACC \cite{mas_openacc}.  It can run simulations containing over three hundred million grid cells \cite{supermas,supermas2} and exhibits performance scaling to thousands of CPU cores or dozens of GPUs \cite{mas_sts,mas_openacc}.  The MAS code uses a logically rectangular non-uniform staggered spherical grid and finite-difference discretizations with a combination of explicit and implicit time-stepping methods\footnote{\url{https://www.predsci.com/mas}}.  The code is highly memory-bound, with its performance typically proportional to the hardware's memory bandwidth.

\section{From OpenACC to {\tt DO CONCURRENT}}
\label{sec:acc2dc}

Here we describe our implementations of replacing {OpenACC} with DC.  Our strategy results in six separate code versions, starting with the current {OpenACC} MAS implementation.  These versions arise due to several considerations including (not) avoiding code refactoring, (in)adherence to the current Fortran specification, (not) using experimental compiler features, and trade-offs between reducing OpenACC directives and performance.  Among the versions, we were able to achieve a working code with \emph{zero} OpenACC directives (Code 5).

Support for using DC for accelerated computing is in its infancy.  Several vendors have announced future plans to support it, but only NVIDIA and Intel have working implementations, with NVIDIA's HPC SDK's {\tt nvfortran} being the more mature of the two.  As our current GPU implementation of MAS uses OpenACC and is tested with NVIDIA GPUs, we exclusively use the {\tt nvfortran} compiler and NVIDIA GPUs in this work.

The procedure for porting the current OpenACC implementation of MAS (Code 1) to DC is done in several steps.  We first try to avoid base code changes and adhere to the current Fortran 2018 specification, incrementally swapping out OpenACC directives with DC, until we arrive at two versions, one with optimized manual data management (Code 2), and one with automatic data management (Code 3).  We then use the compiler's Fortran 202X preview features to further reduce the number of OpenACC directives and only use automatic data management (Code 4).   By using special compiler flags, alternate code launch scripts, and minimal base code changes, we are able to make a working version that has \emph{zero} OpenACC directives (Code 5).  To improve performance, we also create a version of Code 5 with the OpenACC manual data management directives added back in (Code 6).

In Table~\ref{t:codes} we summarize all six code versions that will be discussed in this section.

\begin{table}[htb]
\centering
\caption{Summary of all MAS code versions developed and tested.\label{t:codes}}
\arrayrulecolor{black}
\begin{tabular}{|l|l|c|c|} 
\hline
\rowcolor{LightCyan}
\begin{tabular}[c]{@{}c@{}}Code\\Version\end{tabular} & \begin{tabular}[l]{@{}l@{}}Code description and\\{\tt nvfortran} GPU compiler flags\end{tabular} & \begin{tabular}[c]{@{}c@{}}Total\\Lines\end{tabular} & \begin{tabular}[c]{@{}c@{}}{\tt \$acc}\\Lines\end{tabular} \\ 
\hline
\hline
\rowcolor{Gray2}
0: CPU & \begin{tabular}[c]{@{}l@{}}Original CPU-only version\end{tabular} & 69874 & $\varnothing$ \\ 
\hline
1: A & \begin{tabular}[c]{@{}l@{}}Original OpenACC \\implementation\\{\tt -acc=gpu -gpu=cc80}\end{tabular} & 73865 & 1458 \\ 
\arrayrulecolor[rgb]{0.753,0.753,0.753}\hline
2: AD & \begin{tabular}[c]{@{}l@{}}OpenACC for DC-incompatible\\loops and data management,\\DC for remaining loops\\{\tt -acc=gpu -stdpar=gpu}\\{\tt -gpu=cc80,nomanaged}\end{tabular} & 71661 & 540 \\ 
\hline
3: ADU & \begin{tabular}[c]{@{}l@{}}OpenACC for DC-incompatible\\loops, DC for remaining loops,\\Unified memory\\{\tt -acc=gpu -stdpar=gpu}\\{\tt -gpu=cc80,managed}\end{tabular} & 71269 & 162 \\ 
\hline
4: AD2XU & \begin{tabular}[c]{@{}l@{}}OpenACC for for functionality\\, DC2X for remaining loops,\\Unified memory\\{\tt -acc=gpu -stdpar=gpu}\\{\tt -gpu=cc80,managed}\end{tabular} & 70868 & 55 \\ 
\hline
5: D2XU & \begin{tabular}[c]{@{}l@{}}DC2X for all loops,\\some code modifications,\\Unified memory\\{\tt -stdpar=gpu -gpu=cc80}\\{\tt -Minline=reshape,name:s2c,}\\{\tt boost,interp,c2s,sv2cv}\end{tabular} & 68994 & {\color{blue} \large \bf $\varnothing$} \\ 
\hline
6: D2XAd & \begin{tabular}[c]{@{}l@{}}DC2X for all loops,\\some code modifications,\\OpenACC for data management\\{\tt -acc=gpu -stdpar=gpu}\\{\tt -gpu=cc80,nomanaged}\\{\tt -Minline=reshape,name:s2c,}\\{\tt boost,interp,c2s,sv2cv}\end{tabular} & 71623 & 277 \\
\arrayrulecolor{black}\hline
\end{tabular}
\end{table}

\subsection{Code 1 [A]: Current OpenACC Implementation}
\label{sec:code1}

We start with the current GPU production branch of MAS.  This is an extension of the code developed in Ref.~\cite{mas_openacc}, with added {OpenACC} directives to accelerate the full set of physics models used in the CORHEL software suite.  It is fully portable with CPU compilers as it does not use any OpenACC API calls.  It has 73,842 lines of code including 1,458 {\tt !\$acc} directive comments.  The distribution of OpenACC directive types is summarized in Table.~\ref{t:acc}.
\begin{table}[htb]
\centering
\caption{OpenACC directives in original \\GPU branch of MAS (Code 1).\label{t:acc}}
\begin{tabular}{|l|c|} 
\hline
OpenACC directive type & \# of lines \\
\hline
{\tt parallel}, {\tt loop} & 997 \\
\begin{tabular}[l]{@{}l@{}}data management:\\\qquad{\tt enter}, {\tt exit}, {\tt update},\\\qquad{\tt host\_data}, {\tt declare}\end{tabular} & 320 \\
{\tt atomic} & 34 \\
{\tt routine} & 12 \\
{\tt kernels} & 6 \\
{\tt wait} & 6 \\
{\tt set device\_num} & 1 \\
\begin{tabular}[l]{@{}l@{}}continuation lines ({\tt!\$acc\&})\\\qquad (spread across all directive types)\end{tabular} & 82\\
\hline
Total & 1458 \\
\hline
\end{tabular}
\end{table}

\subsection{Code 2 [AD]: OpenACC for DC-incompatible loops and data management, DC for remaining loops)}
\label{sec:code2}

In this first pass at replacing OpenACC with DC, we strive to maintain as much performance and portability as possible, with as few code changes as possible (other than converting {\tt do} loops into DC).  

A typical nested {\tt do} loop with OpenACC directives as used in the MAS is shown in Listing~\ref{eq:do}.  The same loop using DC is shown in Listing~\ref{eq:dc}. Besides eliminating OpenACC directives, the DC loop is also much more compact, reducing the number of lines of code overall.
\begin{lstlisting}[language={Fortran}, caption={Standard {\tt do} loop with OpenACC in the MAS code}, label={eq:do}]
!$acc parallel default(present)
!$acc loop collapse(3)
do k=1,n3
  do j=1,n2
    do i=1,n1
      Computation using array(i,j,k)
    enddo
  enddo
enddo
!$acc end parallel
\end{lstlisting}
\begin{lstlisting}[language={Fortran}, caption={Equivalent loop to Listing~\ref{eq:do} using DC}, label={eq:dc}]
do concurrent (k=1,n3,j=1:n2,i=1:n1)
  Computation using array(i,j,k)
enddo
\end{lstlisting}

To ensure that the modified code will still work for CPU-only runs with any compiler vendor/version that currently works with the code, we adhere to the Fortran 2018 standard.  This excludes the use of the upcoming Fortran 202X  {\tt reduce} clause on DC loops, leaving all reduction loops to use OpenACC.  Also, since some major compilers do not yet support DC affinity clauses such as {\tt private} and {\tt shared} (and some are good at detecting proper affinity automatically), we do not make use of these clauses.  Array reductions in the current MAS code use OpenACC {\tt atomic} directives to allow for full parallelization as shown in Listing~\ref{l:ar1}.  
\begin{lstlisting}[language={Fortran}, caption={Example of array reduction loop in MAS}, label={l:ar1}]
!$acc parallel default(present)
!$acc loop collapse(2) 
do j=1,n2
  do i=1,n1
!$acc atomic update
    sum0(i)=sum0(i)+array(i,j)*...
  enddo
enddo
!$acc end parallel
\end{lstlisting}
In order to use DC for the loop and remain within specification, we would need to rewrite the loops (as an outer parallel DC loop, with an inner sequential loop).  Since at this point we want to avoid as many code changes as possible, and such a change may affect performance, we also leave the array reduction loops as OpenACC loops.

Some of our parallel loops contain calls to functions/routines.  In OpenACC, these must be declared at a specified level of parallelism using the {\tt routine} directive.  The Fortran specification requires that function/routine called in a DC loop must be `pure'.  Those within our loops are pure, but the current NVIDIA compiler does not yet support them in DC GPU offload loops without leaving in the OpenACC {\tt routine} directives.

Another issue we faced is the use of OpenACC's {\tt kernels} construct.  While {\tt kernels} can be used on regular {\tt do} loops, it can also be used to contain Fortran array-syntax operations and Fortran intrinsics such as {\tt MINVAL}.  In these cases, in order to use DC, we would have to expand the operations into explicit loops.  In our updated port of MAS to use OpenACC, we have already done this in most cases.  In this DC version, we want to avoid additional code changes, so for now we leave in the remaining few {\tt kernels} regions.

There are two features of OpenACC that are not present in DC which could lead to possible performance losses.  The first is `kernel fusion'.  In OpenACC, one can have multiple data independent loops in a {\tt parallel} region, which the compiler can then compile into a single GPU compute kernel.  When converting these loops into DC loops, the compiler has to create separate GPU kernels for each loop.  Since kernels have launch overheads, this can lower performance (especially for frequently called small kernels).  The second feature is asynchronous kernel launches.  In OpenACC, one can use the {\tt async} clause to launch a kernel but allow the code to continue (where it can compute things on the CPU, and/or launch other asynchronous kernels).  DC currently has no way to tell the compiler to use an asynchronous kernel launch. Therefore, if a code relies on heavy use of {\tt async}, using DC may reduce performance.  In the case of the MAS code, we do not make heavy use of {\tt async}, so we do not expect this to be too much of a problem.

From our previous experience in porting OpenACC to DC\cite{stdpardc}, we learned that using manual data management yields better performance than relying on unified managed memory (UM) capabilities.  Unified managed memory is an NVIDIA feature that will automatically page data to and from the GPU and CPU for GPU-accelerated codes\footnote{\url{https://www.pgroup.com/blogs/posts/openacc-unified-memory.htm}}.  This eases the burden of the programmer by not requiring manual management of the data movement.  However, using UM has some drawbacks, such as possible performance degradation and using too much GPU memory.  While the performance drop from using UM can be quite small, when used with CUDA-aware MPI calls (as done in MAS), they can be higher.  Also, UM is not part of either the Fortran or OpenACC specifications.  For these reasons, we do not use UM in this version (AD), leaving the OpenACC data management directives intact.

In the end, for this version (AD) of the code, we have reduced the number of OpenACC directives from 1458 to 540, an almost three-fold reduction.

\subsection{Code 3 [ADU]: OpenACC for DC-incompatible loops, DC for remaining loops, Unified memory}
\label{sec:code3}

For this version (ADU) of the code, we substantially reduce the number of OpenACC directives by using unified managed memory.  As discussed above, UM is not part of the Fortran or OpenACC specification, and without manual data movement, each OpenACC and DC kernel would default to GPU-CPU data migration, causing a huge performance drop.  Due to this, multiple vendors have created a unified memory management system (UM for NVIDIA, Unified Shared Memory for Intel OneAPI\footnote{\url{https://www.intel.com/content/www/us/en/developer/articles/code-sample/dpcpp-usm-code-sample.html}}, and Smart Access Memory for AMD\footnote{\url{https://www.amd.com/en/technologies/smart-access-memory}}).  Since unified memory management systems are ubiquitous, we feel relying on UM is an acceptable decrease in portability for this version (ADU).  

In removing the OpenACC data directives, we found that we could not remove all of them.  Specifically, a {\tt declare} (and a subsequent {\tt update}) directive was still needed for a data element that is used in a function called inside a GPU kernel region.  Additionally, we had to leave in some {\tt enter} and {\tt exit} data movement for derived types.  This was because we are still using OpenACC for reduction loops, which utilize the {\tt default(present)} clause.  This tells the compiler that all data is already on the device, and is added to avoid programming performance errors since it will fail for any data not on the device.  However, when using UM, even though the arrays within the derived type are paged to the GPU, the derived type structure itself is not, as it is static data (not an allocatable or automatic array which is required for UM paging).  Therefore, since we wish to retain the use of {\tt default(present)}, we have to manually place the structures on the GPU.

This version (ADU) of the code further reduced the number of OpenACC directives by over a factor of three from 540 to 162.  Since the only change in this code was the removal of data directives, its performance is equivalent to running Code 2 (AD) with UM enabled (in which case all data directives are ignored).  Thus, this version (ADU) can be used as a test of the effect UM alone has on performance. 

\subsection{Code 4 [AD2XU]: OpenACC for functionality, DC2X for all loops, Unified memory}
\label{sec:code4}

To see how much further we can reduce the number of OpenACC directives, in this version (AD2XU) of the code we allow ourselves to take advantage of NVIDIA's Fortran 202X preview implementation of the {\tt reduce} clause on DC loops.  Using {\tt reduce} breaks the portability of this version (AD2XU), making it only currently work with the {\tt nvfortran} compiler (even on the CPU).   However, this clause is expected to become part of the Fortran standard within a year, so its use will eventually be portable (as soon as all major compilers recognize it).   

Using {\tt reduce} allows us to convert all scalar OpenACC reduction loops into DC, but for array reductions, direct use of {\tt reduce} is not yet supported in {\tt nvfortran}.  However, due to the underlying compiler mechanisms, we found that we can use OpenACC {\tt atomic} directives within DC loops, in the same manner we have been using them in the OpenACC loops.  Therefore, we have also converted all array reduction OpenACC loops into DC (while retaining the {\tt atomic} directives within them) as shown in Listing~\ref{l:ar2}.
\begin{lstlisting}[language={Fortran}, caption={DC version of array reduction from Listing~\ref{l:ar1}}, label={l:ar2}]
do concurrent (j=1:n2,i=1:n1)
!$acc atomic update
  sum0(i)=sum0(i)+array(i,j)*...
enddo
\end{lstlisting}
We also were able to convert the few non-reduction loops that used {\tt atomic} directives into DC.  After this change, all loops using the derived types are now DC loops, so some of the few remaining data directives were also able to be removed.

This version (AD2XU) of the code further reduced the number of OpenACC directives by more than another factor of three from 162 to 55.   The remaining OpenACC directives consist of {\tt atomic}, {\tt declare}, {\tt update}, {\tt set device\_num}, {\tt routine}, and {\tt kernels}.

\subsection{Code 5 [D2XU]: DC2X for all loops, some code modifications, Unified memory}
\label{sec:code5}

Here, we allow ourselves to make use of {\tt nvfortran}-specific compiler flags and some modifications to the base code.   This is an attempt to see what is needed to currently achieve an `ideal' code (in terms of eliminating OpenACC directives).  

As a first step, we eliminated the few remaining {\tt kernels} directives by expanding the Fortran intrinsic functions being used into explicit DC reduction loops.  We note that for codes that have a lot of {\tt kernels} use, this step could be quite involved.

Next, we re-wrote all array reduction loops, by flipping the loop order and using a standard DC on the outer loop, with a DC reduction on the inner loop as shown in Listing~\ref{l:ar3}.
\begin{lstlisting}[language={Fortran}, caption={Modified array reduction from Listing~\ref{l:ar2}, allowing removal of the {\tt atomic} directive}, label={l:ar3}]
do concurrent (i=1:n1)
  tmp=0.
  do concurrent (j=1:n2) reduce(+:tmp)
    tmp=tmp+array(i,j)*...
  enddo
  sum0(i)=tmp
enddo
\end{lstlisting}
This eliminated the need for the OpenACC {\tt atomic} directives in these loops.  We could have also replaced the inner loop with a serial {\tt do} loop or Fortran intrinsic (e.g. {\tt SUM}).  Indeed, the {\tt nvfortran} compiler output shows that it serializes the inner DC reduction loop (as it is probably faster to do that than launch a multitude of tiny kernels).  Small code modifications were also used in a couple of places to allow the removal of the last few {\tt atomic} directives.

A remaining set of OpenACC directives are the {\tt routine} directives used to declare functions and routines that are called within a GPU kernel loop.  Since all such functions in MAS are `pure', they should be supported within DC loops, and further development of the {\tt nvfortran} compiler should remove the need for the {\tt routine} directives.  In the meantime, we use a compiler flag {\tt -Minline,name} to specify that those routines need to be in-lined.  For one of the routines, we also needed to specify the {\tt reshape} inline option.   The compiler refused to inline one of the routines, forcing us to manually inline it.  Luckily, it was only one routine and not used very many times, but in other codes, this step could be prohibitive.  This step also allowed us to remove the {\tt declare} directive for the data elements used within the in-lined functions.

The last OpenACC directive left is the {\tt set device\_num} used to select the GPU device based on the local MPI rank in multi-GPU runs.  In order to remove this directive, we modify how we launch the code.  Instead of launching the code directly (e.g. {\tt mpirun ... ./mas ...}), we launch a bash script (shown in Listing~\ref{l:script}) that 
 uses an MPI run time environment variable to set an NVIDIA environment variable, allowing the MPI process to only see the correct GPU device.  The script is invoked by running {\tt mpirun ... ./launch.sh ./mas ...}.
\begin{lstlisting}[language={Bash}, caption={Bash script ({\tt launch.sh}) used to launch Codes 5 (D2XU) and 6 (D2XAd).}, label={l:script}]
#!/bin/bash
# Assume 1 GPU per MPI local rank
# Set device for this MPI rank:
export CUDA_VISIBLE_DEVICES="$OMPI_COMM_WORLD_LOCAL_RANK"
# Execute code:
exec $*
\end{lstlisting}
While this example is specific to the OpenMPI library (which is bundled with the NV HPC SDK), similar environment variables exist in other MPI libraries.  

With the previous step completed, we have achieved our goal of obtaining a version of the code with \emph{zero} OpenACC directives that can be run on multiple GPUs.  It also allows us to reduce the size of the code further by removing a series of duplicate routines that were needed in the OpenACC implementation.   A number of routines in MAS are called both in the setup phase and the computational portion of the code.  Since the setup phase contains a very large amount of code but accounts for only a negligible amount of run time, we did not want to port all of the setup code to GPUs.  This caused a problem since the ported routines assume the data is on the GPU.   Now that we are using unified managed memory, we can remove the CPU-only versions of these routines, as the GPU-CPU paging overhead during the small setup phase should not significantly affect the overall performance of the code.

\subsection{Code 6 [D2XAd]: DC2X for all loops, some code modifications, OpenACC for data management}
\label{sec:code6}

As we will see in the next section, the use of UM has a very bad effect on performance for the MAS code.  Therefore, we add one more version of the code to our list.  In this version (D2XAd), our goal is similar to Code 5 (D2XU) in trying to get the absolute lowest number of OpenACC directives possible (using experimental features, code modifications, etc.), but here we also require that the performance is on par with Code 1 (A) and Code 2 (AD).  To do this, we start with Code 5 (D2XU) and put back in all the OpenACC manual data movement directives to allow us to run without using UM (this also required us to put back all the duplicate CPU-only routines mentioned in the previous section).  We then modified the code to use wrapper routines for creating and initializing arrays on the GPU, reducing the number of required data movement directives. The resulting code has 277 OpenACC directives, which is over 5 times fewer than our original Code 1 (A), and almost $50\%$ less than Code 2 (AD), while retaining similar performance (see next section).  

\section{Performance tests}
\label{sec:perf}

\subsection{Test Case}
\label{sec:test}
\begin{figure}[htb]
    \centering
    \includegraphics[width=0.75\columnwidth]{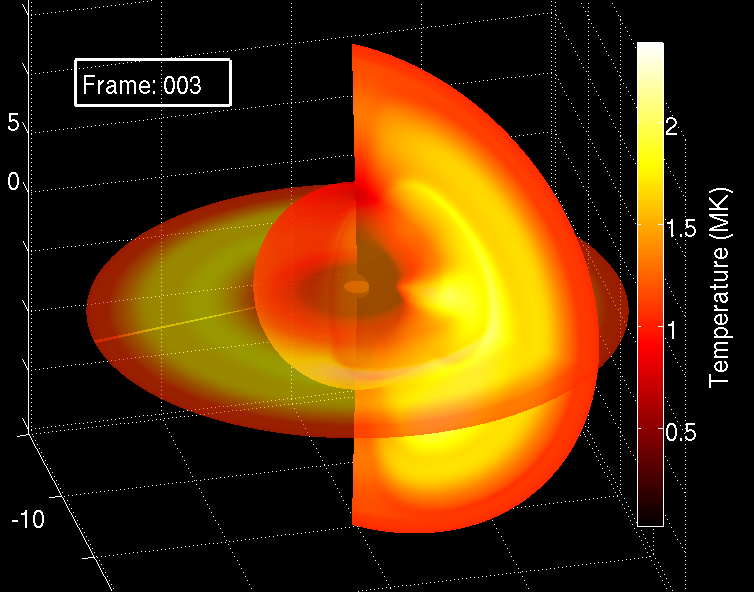}
    \caption{Visualization of the MAS solution for the test case.  Cuts of temperature from the last time step are shown and plotted on a stretched grid. \label{fig:test} }
\end{figure}

To test the performance of our implementations, we use a production quasi-steady-state coronal background simulation from Ref.~\cite{cme1997} that uses a full thermodynamic MHD physics model.  The problem is set to a resolution of 36 million cells and run for the first 24 minutes of its 48 hour physical simulation time.  A visualization of the resulting solution is shown in Fig.~\ref{fig:test}. 
The resolution is chosen to represent a medium-sized case that can also fit into the memory of a single NVIDIA A100 (40GB) GPU.  For all test runs, the solutions were validated against that of the original code to within solver tolerances.

\subsection{Hardware and Software Environment}
\label{sec:compenv}
We test the code versions using a single 8xGPU node of NCSA's Delta\footnote{\url{https://delta.ncsa.illinois.edu}} supercomputer.  The node is dual-socket, with two AMD EPYC Milan 7763 CPUs and a total of eight NVlink-connected NVIDIA A100 (40GB) GPUs, each of which has a peak theoretical memory bandwidth of 1,555 GB/s.   We compile the codes with the NVIDIA HPC SDK's {\tt nvfortran} compiler version 22.11 and the OpenMPI library version 3.1.3 (we did not use OpenMPIv4 with UCX as its performance and compatibility was not on par with OpenMPIv3). 
For our baseline CPU tests, we use the dual-socket AMD EPYC 7742 CPU nodes on SDSC's Expanse\footnote{\url{https://www.sdsc.edu/services/hpc/expanse}} supercomputer, each having a maximum theoretical memory bandwidth of 381.4 GiB/s (409.5 GB/s).  We use GCC's {\tt gfortran} compiler version 10.2.0 with the OpenMPI MPI library version 4.0.4.

\subsection{Performance Results}
\label{sec:performance}

Here we test the performance of each of the six code versions described in Sec.~\ref{sec:acc2dc}.  First, we run Code 1 (A) and Code 2 (AD) on the dual-socket AMD EPYC CPU nodes on Expanse.  Besides giving a baseline performance, this also checks that using DC does not negatively affect portability and performance (at least for the specification-compliant version).  The results are shown in Table.~\ref{t:cputime} and show that the DC version of the code runs equivalently to the original code on the CPUs.
\begin{table}[htb]
\centering
\caption{Wall clock time (in minutes) for the test problem run on dual-socket AMD EPYC 7742 CPU nodes.\label{t:cputime}}
\begin{tabular}{|c|r|r|} 
\hline
\# Nodes & Code 1 (A) & Code 2 (AD) \\
\hline
1 & 725.54 & 725.53\\
8 & 79.58 & 79.64\\
\hline
\end{tabular}
\end{table}

\begin{figure}[htb]
    \centering
    \includegraphics[width=\columnwidth]{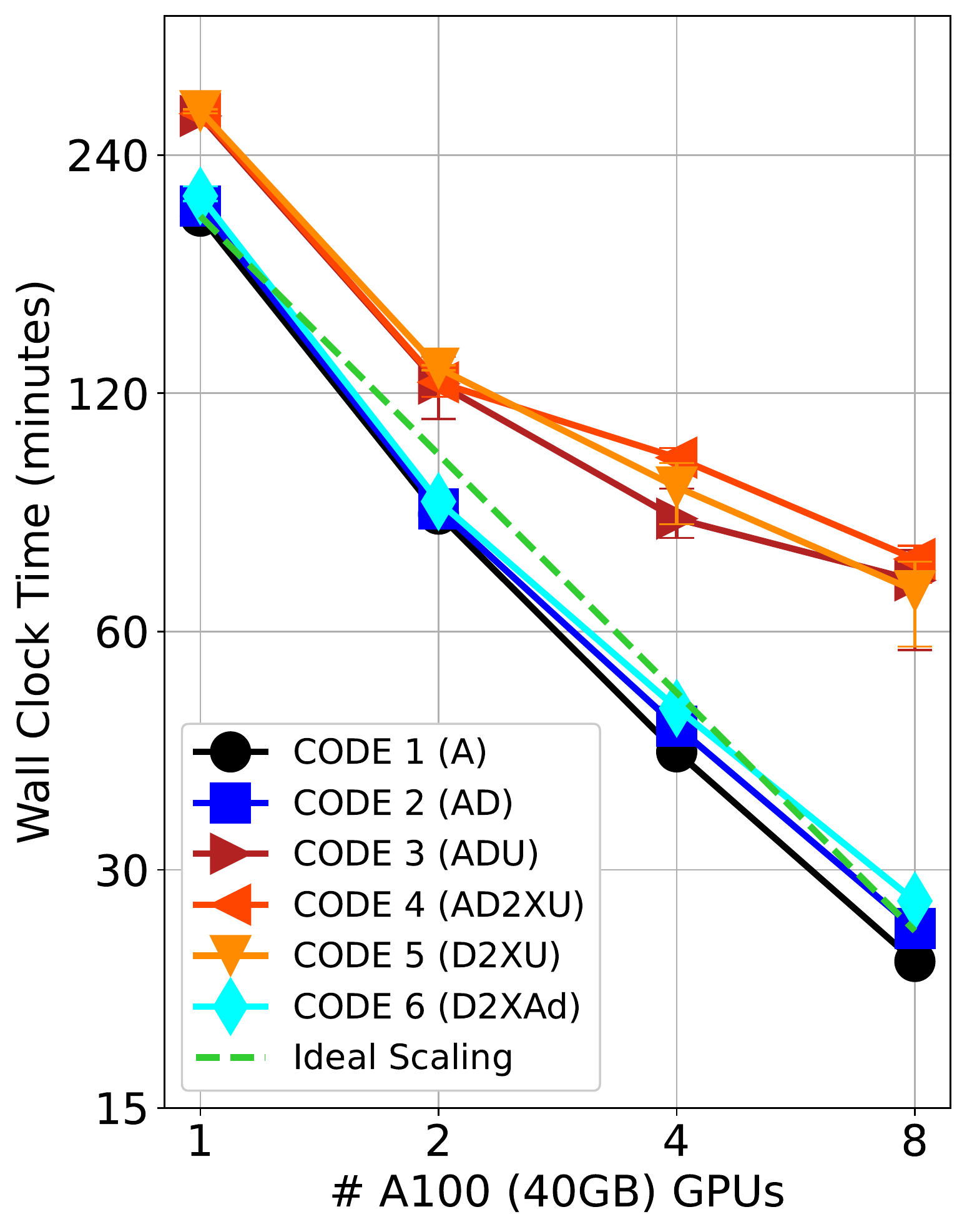}
    \caption{Wall clock time for the test problem run on an 8xA100 (40GB) GPU node.  For each result, the average of three runs is shown, with the minimum and maximum time runs shown as an error bar.\label{fig:scaling} }
\end{figure}

\begin{figure}[htb]
    \centering
    \includegraphics[width=\columnwidth]{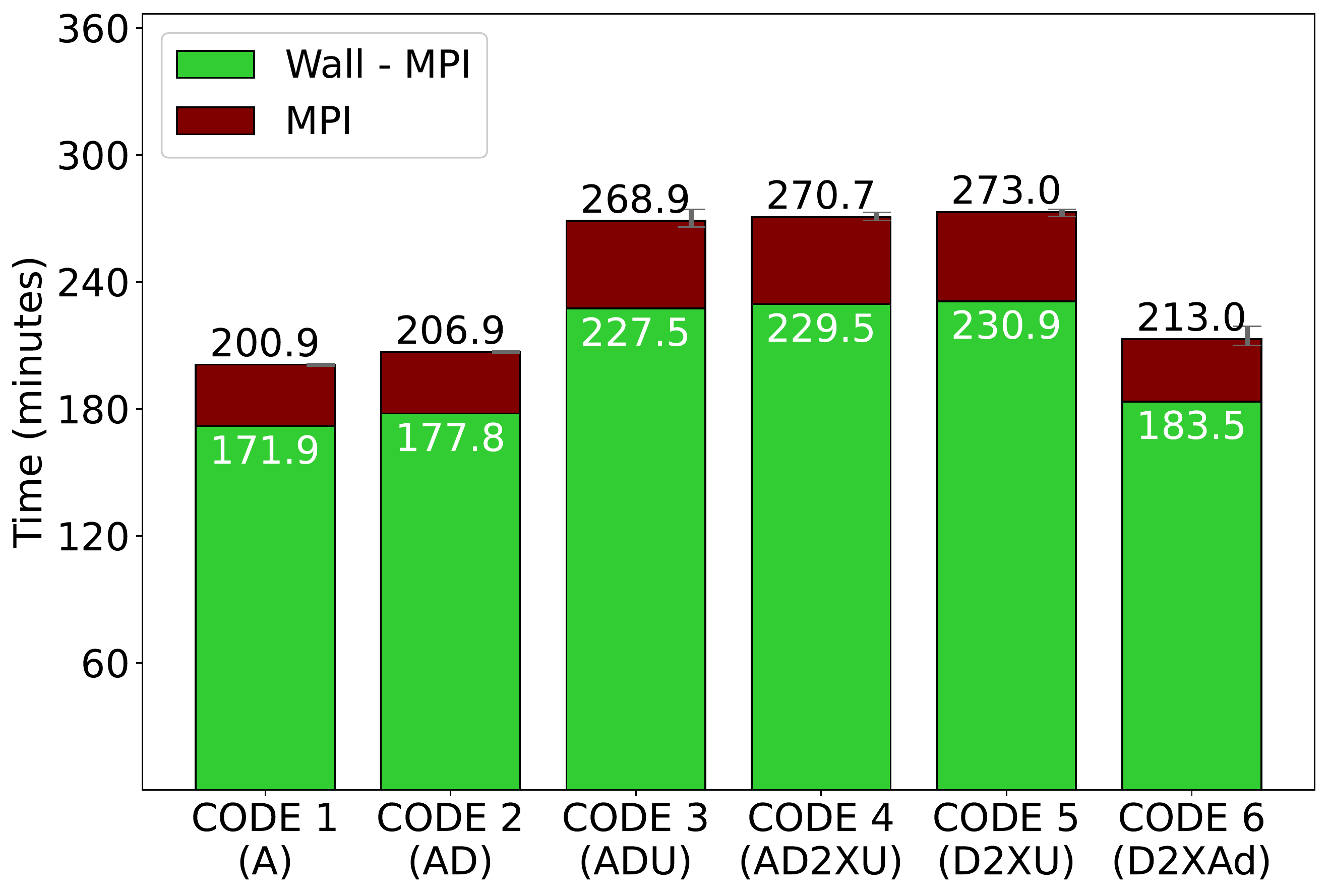}
    \includegraphics[width=\columnwidth]{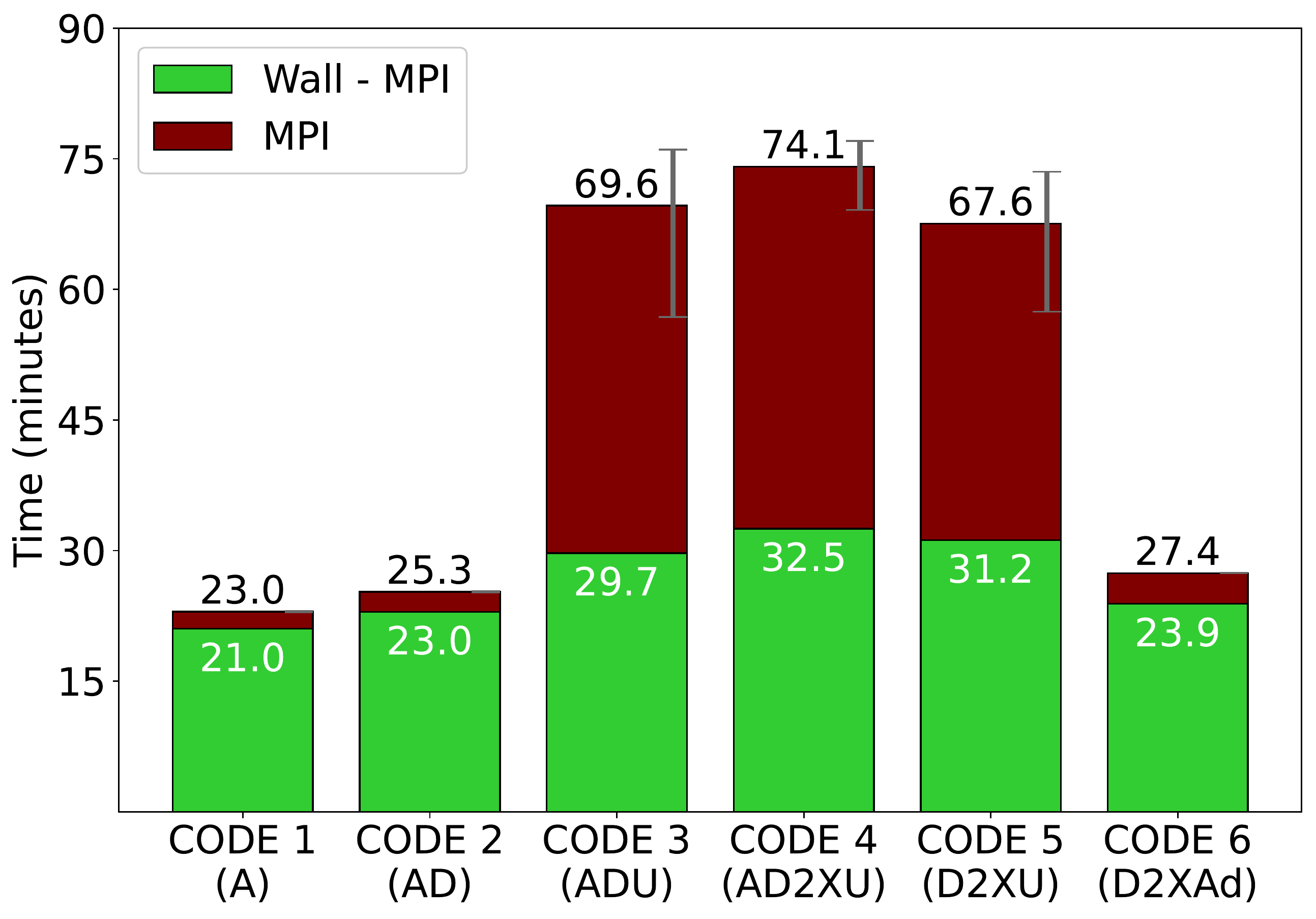}
    \caption{Run times for the test problem on 1 (top) and 8 (bottom) A100 (40GB) GPUs.  For each result, the average of three runs is shown, with the minimum and maximum wall clock times shown as an error bar.  The MPI time (including all MPI calls, buffer initialization/loading/unloading, and MPI waiting caused by load imbalance) is shown in maroon, while the remainder of the wall clock time is shown in green.\label{fig:bars}}
\end{figure}

\begin{figure*}[t]
    \centering
    \includegraphics[width=\textwidth]{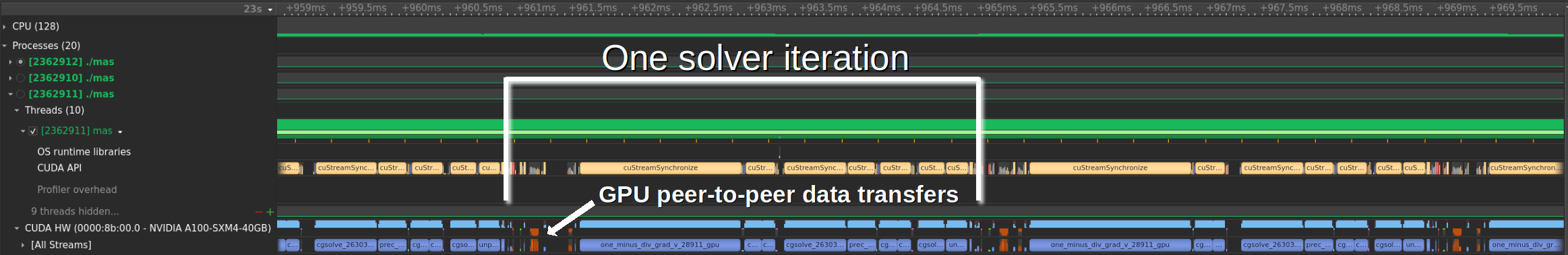} \newline
    \vspace{0.1pt}
    \includegraphics[width=\textwidth]{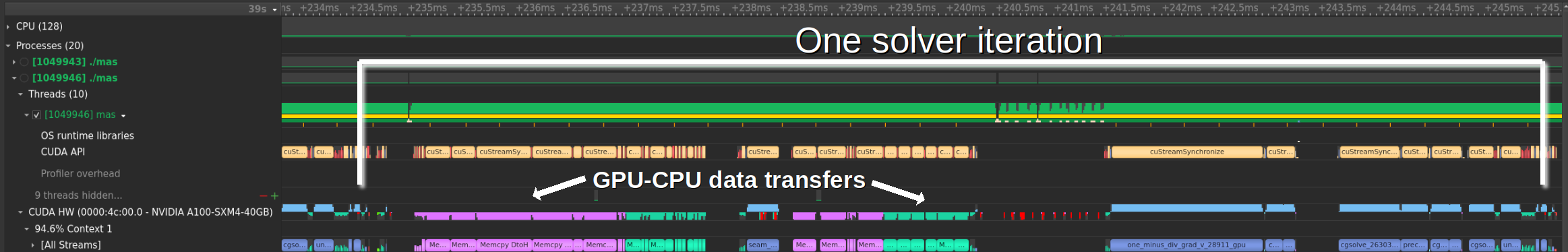}
    \caption{NVIDIA NSIGHT Systems time profile of viscosity solver iterations in MAS using OpenACC manual memory management (top) and unified managed memory (bottom) for Code 1 (A) on 8 A100 GPUs. We see that the manually managed memory results in GPU peer-to-peer data transfers within the MPI halo exchanges, while the unified memory performs multiple CPU-GPU transfers, leading to a slowdown. This, along with more overhead, makes computing a solver iteration three times slower with unified memory management than with manual memory management.\label{fig:mmvsum} }
\end{figure*}

The timings for all six versions of the code run from one to eight NVIDIA A100 (40GB) GPUs on a Delta GPU node are shown in Fig.~\ref{fig:scaling}.
We see that Codes 1 (A), 2 (AD), and 6 (D2XAd) exhibit `super' scaling at first, and then the scaling dips below the ideal rate.  However, due to the initial super scaling, all these codes exhibit better than or close to ideal scaling at 8 GPUs.  We also see that Codes 2 (AD) and 6 (D2XAd), which use DC, are both somewhat slower than Code 1 (A).  Possible reasons for this are kernel fission, loss of asynchronous kernels, and different compiler offload parameters between the OpenACC and DC kernels.  Code 6 (D2XAd) is also seen to be a bit slower than Code 2 (AD), which is likely due to additional array initialization kernels in the wrapper routines in places where the original code did not initialize the arrays to zero. Codes 3 (ADU), 4 (AD2XU), and 5 (D2XU) have greatly reduced performance and scaling.  In Fig.~\ref{fig:bars}, we show the results for 1 and 8 GPUs, highlighting the MPI overhead time.
The MPI time is greatly increased in the codes that use UM, and the non-MPI time is increased as well (but to a much smaller degree).  All the codes that exhibit worse performance have similar timings, and all use UM.  Since Code 3 (ADU) is equivalent to running Code 2 (AD) with UM enabled, these results indicate that the UM is the cause of the performance drop, not DC.  We confirmed this by running Code 1 (A) and Code 2 (AD) with UM and got similar timings to Code 3 (ADU).   To explore why UM is causing such a large drop in performance, we ran the NVIDIA NSIGHT Systems profiler for the 8-GPU run of Code 1 (A) with both manual memory management and UM. The results are shown in Fig.~\ref{fig:mmvsum}.
We see that the manually managed memory allows for GPU peer-to-peer data transfers in the MPI halo exchanges, while the UM performs multiple CPU-GPU transfers, leading to a slowdown. The UM run also had more overhead as seen in the larger gaps between kernel launches.  Overall, the manual memory management run completes almost three full iterations in the same time it takes the UM run to complete one.  The developers at NVIDIA have recognized this problem, and are working on a way to resolve it\footnote{Jeff Larkin, NVIDIA Corp. personal correspondence}.

\section{Summary and outlook}
\label{sec_summary}
Out of the six code versions we tested, we highlight three.  First, Code 1 (A, our original OpenACC code) is the best performing version.  This is due to manual memory management, and OpenACC's ability to have asynchronous kernel launches, and kernel fusion.  
The second code we highlight is Code 5 (D2XU).  This code shows the potential of standard languages as it can run on multiple GPUs without using a single OpenACC directive.  Although it suffers from performance degradation due to UM, we expect this to be a temporary setback which will be solved with further development of the compilers, MPI libraries, and system integration improvements.  Code 5 (D2XU) also required us to inline some functions by hand, use inlining compiler flags, and use an alternative method for launching the code with MPI.  Some of these requirements are expected to be avoided with future compiler releases as well.
The last code we highlight is Code 2 (AD).  It adheres to the current Fortran specification, has performance nearly as good as Code 1 (A), can still compile with all major CPU compilers, and has greatly reduced the number of OpenACC directives from 1458 to 540.  It shows that the performance of the DC loops are competitive with OpenACC as long as one uses manual GPU-CPU data movement.  Adding in some of the Code 5 (D2XU) and Code 6 (D2XAd) modifications (as well as others) can bring the number of OpenACC directives even lower, and such a modified version has become our new production GPU version.  

With further development and cross-vendor support, we hope to eventually have a single code base capable of running on multiple vendors' accelerator hardware without the need for directives at all.  This will be extremely valuable for maintaining a high-performance, portable, accelerated code, while also allowing domain scientists to develop it in standard Fortran.  Although directives can be more straightforward to use than low-level APIs, they still require domain scientists to learn an unfamiliar syntax (and how to properly utilize it), hindering productivity.  With a standard language approach, they can focus on their implementations, comfortable in the language (with minor adjustments) they have been using for decades.

In the meantime, for our codes, we expect to need some amount of OpenACC support for years to come.  We highly encourage vendors to maintain and/or add such support (as it is often used instead of OpenMP for GPU applications), until the standard languages are ready for full cross-vendor, cross-platform use.

\section*{Acknowledgments}
This work is supported by NSF awards AGS 2028154 and ICER 1854790, and NASA grants 80NSSC20K1582 and 80NSSC22K0893.  It also utilized the Delta system at NCSA and the Expanse system at SDSC through ACCESS allocation TG-MCA03S014.  We thank Jeff Larkin at NVIDIA Corp. for his useful insights and the anonymous reviewers for their suggestions.

\bibliographystyle{IEEEtran}
\bibliography{ref.bib}

\end{document}